\documentclass[review,preprint,authoryear]{elsarticle}
\usepackage[a4paper]{geometry}
\usepackage[utf8]{inputenc}
\usepackage{a4wide}
\usepackage{amssymb,mathtools,amsthm}
\usepackage{color}
\usepackage{graphicx}
\usepackage{url}
\usepackage{csquotes}
\usepackage{hyperref}
\usepackage{textgreek}
\usepackage{relsize}
\usepackage{enumitem}
\usepackage{array}
\usepackage[separate-uncertainty = true,multi-part-units=single]{siunitx}
\usepackage{tikz}
\usetikzlibrary{decorations.pathreplacing, positioning, calc, arrows, quotes}
\usepackage[capitalise,noabbrev]{cleveref}
\usepackage[textsize=tiny]{todonotes}
\usepackage[boxruled]{algorithm2e}
\newcommand{\motifgaussian}{g_{\text{motif}}}
\newcommand{\motifgaussianparameters}{\Theta_{\text{motif}}}

\newcommand{\vb}[1]{\ensuremath{#1}}

\renewcommand{\vee}{v}
\newcommand{\EMotifIm}{E_\text{motif}^\text{image}}
\newcommand{\EMotifAtm}{E_\text{motif}^\text{atoms}}
\newcommand{\unitvec}{v}
\newcommand{\ProjUCEuclid}{P_\text{EC}}
\newcommand{\ProjUCCrystal}{P_\text{CC}}
\newcommand{\ProjEuclidToUCCrystal}{P_{\text{E{\smaller\textrightarrow}C}}}
\newcommand{\CoordEuclidToCrystal}{T_{\text{E{\smaller\textrightarrow}C}}}
\newcommand{\CoordCrystalToEuclid}{T_{\text{C{\smaller\textrightarrow}E}}}

\DeclarePairedDelimiter\floor{\lfloor}{\rfloor}

\providecommand{\Abs}[1]{\left\lVert#1\right\rVert}
\providecommand{\brac}[1]{\left(#1\right)}
\newcommand{\setof}[2]{\left\{{#1}\,:\,{#2}\right\}}
\DeclareMathOperator{\dx}{d\mathit{x}}
\def\Xint#1{\mathchoice
      {\XXint\displaystyle\textstyle{#1}}%
      {\XXint\textstyle\scriptstyle{#1}}%
      {\XXint\scriptstyle\scriptscriptstyle{#1}}%
      {\XXint\scriptscriptstyle\scriptscriptstyle{#1}}%
      \!\int}
\def\XXint#1#2#3{{\setbox0=\hbox{$#1{#2#3}{\int}$}
      \vcenter{\hbox{$#2#3$}}\kern-.5\wd0}}

\def\dashint{\Xint-}
\newlength{\imageheight}
\newlength{\imgwidth}

\definecolor{ucvecgreen}{RGB}{  0  132 0}

\title{Direct Motif Extraction from High Resolution Crystalline STEM Images}
\author[1]{Amel Shamseldeen Ali Alhasan}
\author[2]{Siyuan Zhang}
\author[1]{Benjamin Berkels}
\address[1]{AICES Graduate School, RWTH Aachen University, Schinkelstr. 2, 52062 Aachen, Germany}
\address[2]{Max-Planck-Institut für Eisenforschung, Max-Planck-Straße 1, 40237 Düsseldorf, Germany}
\begin{document}
\begin{frontmatter}
\begin{abstract}
    During the last decade, automatic data analysis methods concerning different aspects of crystal analysis have been developed, e.g., unsupervised primitive unit cell extraction and automated crystal distortion and defects detection. %
    However, an automatic, unsupervised motif extraction method is still not widely available yet.%

    Here, we propose and demonstrate a novel method for the automatic motif extraction in real space from crystalline images based on a variational approach involving the unit cell projection operator. Due to the non-convex nature of the resulting minimization problem, a multi-stage algorithm is used. First, we determine the primitive unit cell in form of two lattice vectors. Second, a motif image is estimated using the unit cell information. Finally, the motif is determined in terms of atom positions inside the unit cell.
    The method was tested on various synthetic and experimental HAADF STEM images.  %
    The results are a representation of the motif in form of an image, atomic positions, primitive unit cell vectors, and a denoised and a modeled reconstruction of the input image.
    The method was applied to extract the primitive cells of complex $\mu$-phase structures Nb\textsubscript{6.4}Co\textsubscript{6.6} and Nb\textsubscript{7}Co\textsubscript{6}, where subtle differences between their interplanar spacings were determined.

\end{abstract}
\end{frontmatter}

\section{Introduction}

Transmission Electron Microscopy (TEM) and Scanning Transmission Electron Microscopy (STEM) of crystalline specimens yield images at atomic resolution and thus an abundant amount of data. Various information can be extracted from this data. For example, the parameters of the crystal lattice, the positions of the atomic columns and the local lattice deformation.%

Crystalline TEM/STEM images are just an example of periodic images.
The extraction of the motif of periodic data is an attractive task in various fields across different dimensions.  
For example, one-dimensional motif extraction \citep{wilson_2008} is a point of interest in economics and finance. For time series, \citep{vahdatpour_2009} proposed a  method for finding the motif in multidimensional data.  
Moreover, various research are concerned with the motif together with its symmetries. This is done very recently for example through machine learning \citep{chen_2022} or more classically using wallpaper groups to also find the symmetries of the extracted motif~\citep{miller_1973}. The method proposed in \citep{liu_2004} considers automatic extraction of the motif and all its wallpaper symmetries in 1-D and 2-D. Wallpaper groups are of high interest for motif extraction in tile and fabric, for example \citep{ngan_pang_yung_2011}  as well as in architecture, e.g., \citep{albert_2015}. 

For crystalline TEM/STEM images, there are variety of tools to extract various local and average properties. The limiting factors here are the spatial resolution and high levels of noise. In this context, an important desirable aspect for efficient extraction is automation. The average lattice parameters are often calculated manually from the Fourier transform of the image. An automatic method for the extraction of the primitive unit cell vectors in real space from high resolution STEM images is proposed by \citep{mevenkamp_berkels_2015}. An advantage of using real space methods is to avoid artifacts introduced by the Fourier transform. In three dimensions, the UnitCell Tools \citep{shi_li_2021} find the lattice parameters from electron diffraction patterns. Meanwhile other research are concerned solely with the crystallographic symmetry like in \citep{tiong_2020}.

On the other hand, various tools are concerned with local parameters. For example, Elsey and Wirth propose a method for the automatic extraction of a deformation tensor to locate crystal defects \citep{elsey_wirth_2014}. 
Software tools that address various (local) properties like PREStem~\citep{han_2022}, Atomap~\citep{nord_2017}, StatSTEM~\citep{debacker_2016}, Ranger~\citep{jones_nellist_2013} and qHAADF~\citep{galindo_kret_sanchez_2007} are available. %
 In addition to the distinction between average and local properties, one can also distinguish between model-based approaches and those based on machine learning. Atomap is an example of a 
model-based software. There, the center of mass and Gaussian fitting are used to find and refine the atomic column positions. Moreover, it extracts different atomic species separately by fitting the atoms with high intensity first and then removing them. %
On the other hand, PREStem is an example of a software that uses machine learning to analyze the atomic positions. %

Instead of extracting a large amount of local information, e.g., positions of all atoms in an image, like many of the tools above, we aim at condensing the information contained in a crystalline image by extracting the unit cell and the motif. In case the image is truly periodic, the position of all atoms in the image follows from this condensed representation.

\section{Overview}
In this paper, we propose a novel method to characterize the unit cell in a given high resolution image of a crystalline specimen acquired using scanning transmission electron microscopy (STEM). The output of the method is the crystal's unit cell and its motif either in form of an image or as the positions of its constituent atoms.

We start off by introducing the geometrical formulation of crystals and then define the unit cell and the primitive unit cell vectors. The term \enquote{primitive} is usually omitted in this work, however, it is to be understood that the vectors and the unit cell are primitive.
The extraction of an initial estimate of those vectors is made following \citep{mevenkamp_berkels_2015},
with a few modifications to that approach. %
From there, the concept of the projection to the unit cell is explained.

\begin{figure}
    \centering
    \begin{tikzpicture}
  [font=\small,
   block/.style ={rectangle, draw=black, thick,
   align=center, rounded corners,
   minimum height=1.5em},
   line/.style ={draw, thick, -latex',shorten >=2pt}
  ]
  \node [matrix,row sep=7mm] {
  \node [block] (input) {\textbf{Input:} Crystalline image and number of atoms per unit cell};\\
  \node [block] (step 1) {Find initial estimate of the unit cell vectors $\vee$, cf. \cref{sec:UnitCellExtr}};\\ 
  \node [block] (step 2) {Find the motif in terms of an image $u$, refine $\vee$, cf. \cref{secmotifimg}};\\ 
  \node [block] (step 3) {Find the constituent atoms of the motif as parameters $\Theta$, cf. \cref{sec:MotifDOFs}};\\
  \node [block] (output) {\textbf{Output:} $\vee, u, \Theta$}; \\
   };
  \begin{scope}[every path/.style=line]
    \path (input) -- (step 1) ;
    \path (step 1) -- (step 2); 
    \path (step 2) -- (step 3);
    \path (step 3) -- (output);
  \end{scope}
\end{tikzpicture}%
    \caption{An overview of the proposed motif extraction framework.}
    \label{fig:flowchart}
\end{figure}
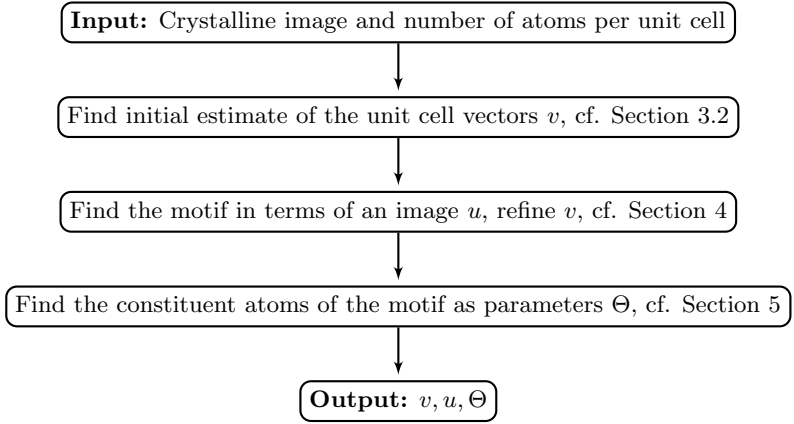

The motif extraction is done in two steps. First, the extraction of the motif in terms of an image. Second, a fitting of two dimensional Gaussian distributions to its atoms. %
In addition to the position of the atoms, this fit includes the estimation of properties such as atom width, $x_1$-$x_2$ correlation (tilt) and pixel intensity.
\cref{fig:flowchart} sketches the entire proposed approach as a flow chart.

Examples of the application of our method to simulated and experimental data are presented. %
Interesting byproducts of this method are denoised and model images of the input image. The denoised reconstruction of the image is used to specify the initial positions of the atoms for the fitting. Moreover, the model image is expected to be useful for further analysis.

Finally, we detail the role of our method in tailoring the plasticity of topologically close-packed phases using Nb-Co $\mu$-phase as an archetypal material \citep{Luo_Xie_Zhang_2023}.

\section{Crystals, primitive unit cells and the unit cell projection}

\subsection{Geometrical formulation of a crystal}

In crystalline materials, the basic unit of the crystal is arranged in a periodic array. This repeated unit can be an ion, molecule, atom, group of atoms, etc.~\citep{ashcroft_mermin_1976,kittel_2004}.
The crystal is defined by two components:
\begin{itemize}
    \item the geometrical arrangement of points set infinitely in space, i.e., the \emph{lattice} $\mathcal{B}$, and
    \item the basic unit attached to each lattice point, i.e., the \emph{motif}. %
\end{itemize}
The lattice is spanned by a number of vectors equal to the dimension of the crystal. In the case of image analysis, the image dimension and therefore the number of these vectors is two. Those vectors need to be linearly independent and define the smallest distance from a lattice point to the next.
Those vectors are called \emph{primitive lattice vectors} $\vee \coloneqq\{\vee_i\}_{i=1}^d$.
Note that these vectors are not unique in the sense that different vectors can describe the same crystal. %
 The (Bravais) lattice corresponding to these vectors is given by the set
\[\mathcal{B} = \setof{\sum_{i=1}^d z_i \vee_i}{z_1, \dots, z_d \in \mathbb{Z}}.\]%
The motif is described by the position of each of its constituent atoms $m_i\in\mathbb{R}^d$ and a value $c_i\in\mathbb{R}$. Thus, the motif is
\[M :=\setof{(m_j,c_j)}{j\in\{1,...,l\}},\]
where $l\in\mathbb{N}$ is the number of atoms in the unit cell.
Putting the two components, i.e., unit cell and motif, together, the crystal can be defined as
\[
    \mathcal{C}
    :=
    \setof{
    \brac{\vb{m}_j + \sum_{i=1}^d z_i \vee_i
    ,
    c_j}}
    {
    z_1, \dots, z_d \in \mathbb{Z}
    ,j\in\{1,...,l\}
    }.
\]

\subsection{Primitive unit cell extraction}
\label{sec:UnitCellExtr}

Let $f:\Omega\to\mathbb{R}$ be an image. Denoting width and height of the image in pixels by $w$ and $h$, respectively, the image domain $\Omega$ can be expressed in pixel coordinates as $\Omega=[0, w-1]\times[0, h-1]$. For the extraction of the primitive unit cell vectors from $f$, we follow the automated three-step method proposed in \citep{mevenkamp_berkels_2015,mevenkamp_2017} based on a real space analysis.

The first step is to find the directions corresponding to high periodicity in the image. For this, the Radon transform is computed. This transform is the projection of the image onto a line for all angles $\delta \in [0^{\circ},180^{\circ})$. To ensure that the integration domain is independent of the projection angle, the image is restricted to the largest circle centered in the rectangular image domain. Furthermore, the projection is normalized by a point-wise division with the Radon transform of the characteristic function of that circle.
From this, the projective standard deviation ($\text{psd}$), i.e., the standard deviation of the projection's intensity along each projection angle $\delta$, is computed via:
\[\text{psd}^2(\delta) = \dashint_{p_\text{min}}^{p_\text{max}} \left(A_\delta(p) - \mu_\delta \right)^2 \,\mathrm{d}p.\]
Here, $A_\delta(p)$ is the intensity at a point $p$ on the projection line defined by the projection angle $\delta$ and $\mu_\delta$ is the average intensity of that line. Note that this approach to compute the $\text{psd}$ slightly differs from how this quantity is computed in \citep{mevenkamp_berkels_2015}.
Projection angles corresponding to high periodicity in the image have a high $\text{psd}$ value and, in particular, are local maxima of the $\text{psd}$. Perpendicular to these angles are the high periodicity directions in the image. Thus, if the projection angle $\delta$ is a local maximum of $\text{psd}(\delta)$, the vector $\hat{e}_{\alpha}=(\sin(\alpha),\cos(\alpha))$ for $\alpha = \delta + \frac{\pi}{2}$ is a high periodicity direction of the image.

The second step is to find the fundamental period $t$ along these directions $\hat{e}_{\alpha}$ to form a set of candidate primitive unit cell vectors. To this end, the local minimizers of the energy $E_{\alpha}$ given by
\begin{equation}
    \label{dirsenr}
    E_\alpha(t) = \int_{\hat\Omega} \left(f(x+t\hat{e}_\alpha) - f(x)\right)^2
        \,
        \mathrm{d}x,
\end{equation}
are candidates for the fundamental period along the direction $\hat{e}_{\alpha}$. Here, $\hat\Omega=[\frac14w,\frac34w-1]\times[\frac14h,\frac34h-1]$ is a restricted integration domain, which ensures that the energy can be evaluated for all $t\in[0,\min(h,w)/4]$%
. The Akaike information criterion (AIC) \citep{akaike_1981} is used to determine the number of clusters $C\in\mathbb{N}$ in the set of energies corresponding to the local minimizers. The minimizers are then clustered using $k$-means clustering into $C$ clusters. Then, a primitive unit cell vector candidate along this direction is the one with the shortest length in the global minimum cluster, i.e., the cluster with the center corresponding to the lowest energy. For more details on the clustering, we refer to \citep{mevenkamp_berkels_2015}.
Then, we get an initial estimate $\{\hat\vee_1,\hat\vee_2\}$ for the primitive unit cell vectors by taking the two shortest non-colinear candidate vectors.

The final step is the refinement of these two vectors by minimization of the energy
    \begin{align*}
        E(\vee_1,\vee_2) &= \sum_{(z_1,z_2) \in \{(1,0),(0,1),(1,1)\}} \int_{\tilde \Omega} \hspace{-0.25em}\left( f(x) - f(x+z_1 \vee_1+z_2 \vee_2) \right)^2 \mathrm{d}x.
    \end{align*}
using the vectors from the previous step as initial guess for the iterative minimization. Here, the integration domain is $\tilde \Omega=[d, w-1-d]\times[d, h-1-d]$ with $d=\max(\Abs{\hat\vee_1}_\infty,\Abs{\hat\vee_2}_\infty, \Abs{\hat\vee_1+\hat\vee_2}_\infty+3)$ given that $\hat\vee_1$ and $\hat\vee_2$ are in pixel coordinates. This ensures that $x+z_1 \vee_1+z_2 \vee_2$ is still in $\Omega$ for $\vee_1$ and $\vee_2$ close to their initial estimate.
For the minimization, we interpret the minimization problem as nonlinear leasts squares problem and solve it numerically using the Gauss--Newton algorithm with adaptive dampening, which ensures a decay of the residual.

\begin{figure}[ht]
\centering
\setlength{\imgwidth}{.4\linewidth}
\begin{tabular}{@{}cc@{}}
    \includegraphics[height=\imgwidth]{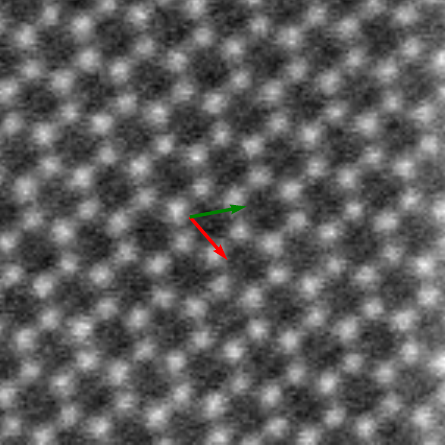}
    &
    \includegraphics[height=\imgwidth]{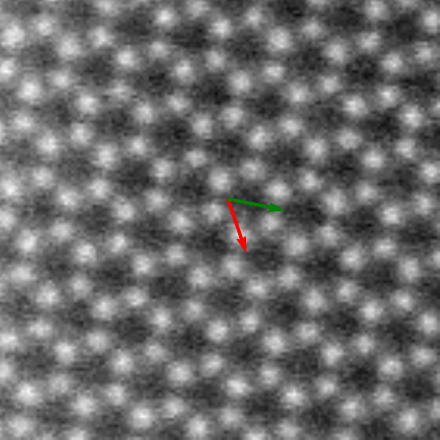}
    \\
    a)
    &
    b)
\end{tabular}
\caption{An example showing the extracted primitive unit cell vectors from two neighboring grains of a magnesium crystal imaged along the [0001] zone axis.}
\label{2grains}
\end{figure}

Our implementation, varied slightly from the method described in \citep{mevenkamp_berkels_2015}. Below, we list these differences, our choice of parameters and assumptions that have not been specified so far:

\begin{itemize}
    \item Our implementation assumes a primitive unit vector to have a length of at least 5 pixels.
    \item Given our choice of $\tilde \Omega$ in Equation (\ref{dirsenr}) described above, the image must at least enclose four unit cells in horizontal and vertical direction for the clustering step to be applicable. %
    \item For the automatic determination of the periodicity directions, a $\text{psd}$ value is considered to be \emph{sufficiently high} if it is bigger than $2.5 \sigma_{\text{psd}}$ where $\sigma_{\text{psd}}$ is the standard deviation of the $\text{psd}$. Local maxima that are not sufficiently high are omitted as they present low periodicity along the corresponding direction.
    \item To numerically compute the Radon transform, the interval $[0^{\circ},180^{\circ})$ is discretized with an angle increment of $0.5^{\circ}$.
    \item Given all unit cell vectors candidates, each corresponding to a minimizer of $E_\alpha$, cf. \eqref{dirsenr}, for a direction $\alpha$, the lower the minima, the better the corresponding minimizer describes the periodicity of the image. Thus, the unit cell vector candidates with a high minimization energy $E_\alpha$ in comparison to the lowest minimization energy are disregarded when choosing the shorted unit cell vectors from the candidates. In other words, the minimization energy is considered as an indicator of the eligibility of the corresponding direction for being the direction of a primitive unit cell vectors.
     \item When selecting $v_1$ and $v_2$ as the shortest vectors, several unit cell vector candidates may have (numerically) the same (shortest) length. If this happens when choosing $\vee_1$, we choose the vector with the smallest angle to the $x_1$-axis. If this happens when choosing $\vee_2$, we choose the vector with smallest angle to $\vee_1$. This way, similar input images should lead to comparable unit cell vectors, while a small angle between the two vectors is preferred. %
\end{itemize}
As mentioned above, the primitive unit cell vectors are not unique. \cref{2grains} exemplifies the effects of our unit cell vector selection strategy. The two images shown in this figure are neighboring grains of a magnesium crystal imaged along the [0001] zone axis. %

\subsection{Projection to the unit cell}
\begin{figure}
\centering
\includegraphics[width=.5\linewidth]{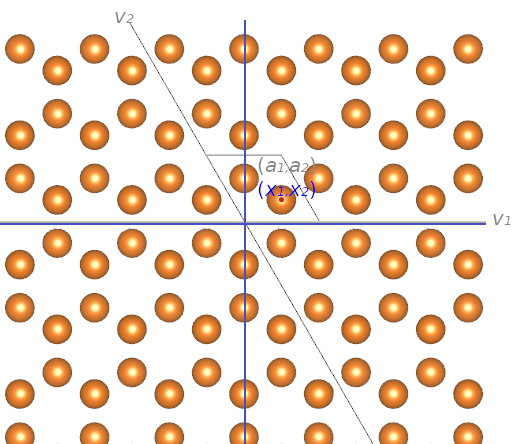}
\caption{Euclidean coordinates $(x_1,x_2)$ and crystal coordinates $(a_1,a_2)$, visualized with VESTA 3 \citep{momma_izoumi_2011,wyckoff_1963}.}
\label{fig:Coords}
\end{figure}

An essential ingredient of the proposed motif extraction approach is the projection to the unit cell. While any position in the image can be naturally described in the Euclidean space with its pixel coordinates, any point in the crystal is naturally described as linear combination of the unit cell vectors, which we call \enquote{crystal space}. \cref{fig:Coords} illustrates the two coordinate systems.
In crystal coordinates, the unit cell is always $[0,1)^2$ and the projection to it is straightforward, i.e., one just has to extract the fractional part of the coordinates. More specifically, the projection is
\[
    \ProjUCCrystal:\mathbb{R}^2\to[0,1)^2,
    a=(a_1,a_2)\mapsto
    \begin{pmatrix}
        a_1 - \floor{a_1}
        \\
        a_2 - \floor{a_2}
    \end{pmatrix}
    \coloneqq
    \begin{pmatrix}
        s
        \\
        t
    \end{pmatrix}
\]
Here, $\floor{\cdot}$ denotes the floor function, i.e., $\floor{x}$ is the largest integer smaller or equal to $x\in\mathbb{R}$.
The matrix with the unit cell vectors $\unitvec=(\unitvec_1$, $\unitvec_2)$ as columns maps points from crystal coordinates to Euclidean coordinates, i.e.,
\[
    \CoordCrystalToEuclid[v]:\mathbb{R}^2 \to \mathbb{R}^2, a\mapsto
\begin{pmatrix}
\unitvec_{1,1}  & \unitvec_{2,1} \\
\unitvec_{1,2}  & \unitvec_{2,2}
\end{pmatrix}
\begin{pmatrix}
    a_1 \\ a_2
\end{pmatrix}
\eqqcolon
\begin{pmatrix}
    x_1 \\ x_2
\end{pmatrix}
\]
Here $\unitvec_{i,j}$ denotes the $j$-component of the unit cell vector $\unitvec_{i}$.
Since the unit cell vectors are non-colinear, the matrix above is invertible and thus also the mapping $\CoordCrystalToEuclid[v]$.
Using $\CoordCrystalToEuclid[v]$, one can construct the projection to the unit cell in Euclidean coordinates by first mapping Euclidean coordinates to crystal coordinates using $\CoordCrystalToEuclid[v]^{-1}\eqqcolon\CoordEuclidToCrystal[v]$, projecting to the unit cell in crystal coordinates using $\ProjUCCrystal$ and finally mapping the projected position back to Euclidean coordinates with $\CoordCrystalToEuclid[v]$.
Thus, in Euclidean coordinates, the projection to the unit cell is
\[\ProjUCEuclid:\mathbb{R}^2\to\setof{s \vb{\unitvec_1} + t \vb{\unitvec_2}}{s,t\in[0,1)^2},x\mapsto(\CoordCrystalToEuclid[v] \circ \ProjUCCrystal \circ \CoordEuclidToCrystal[v]) (\vb{x}).\]

\section{Extraction of the motif image}
\label{secmotifimg}
Given a good initial guess of the unit cell vectors $\unitvec$, the extraction of the motif consists of two steps.
The first step is to find the motif in terms of an image $u$ in crystal space, i.e., $u:[0,1)^2\rightarrow\mathbb{R}$. This is done by minimizing the following energy
\begin{equation}
    \label{motifimageeq}
    \EMotifIm[u,\unitvec] = \int_{\Omega} (f(x) - u(\ProjEuclidToUCCrystal[\unitvec](x)))^2 \dx
\end{equation}
Here, $\ProjEuclidToUCCrystal[\unitvec]$ is the composition of the coordinate transform from Euclidean coordinates to crystal coordinates corresponding to the unit cell vectors $\unitvec$ and the projection to the unit cell in crystal coordinates, i.e.,
\begin{align*}
\ProjEuclidToUCCrystal[\unitvec] :
 \mathbb{R}^2  \rightarrow [0,1)^2,
 \vb{x}\mapsto(\ProjUCCrystal\circ\CoordEuclidToCrystal[v])(\vb{x}) = (s,t).
\end{align*}
At first glance, it may seem more natural to describe the motif image $u$ in Euclidean space, which would mean to use the projection to the unit cell in Euclidean space $\ProjUCEuclid$ instead of $\ProjEuclidToUCCrystal$ in \eqref{motifimageeq}. Discretizing, the unit cell in Euclidean space, i.e., $\setof{s \vb{\unitvec_1} + t \vb{\unitvec_2}}{s,t\in[0,1)^2}$, with a pixel grid and an interpolation that can handle periodicity at the unit cell boundaries, is not straightforward though.
With $u$ described in crystal space, such an interpolation can be done rather easily. Let $U$ be a discretization of $u$ with a regular pixel grid, which is essentially a matrix. Then, one can use standard interpolation on a rectangular domain while extending $U$ with one additional row and column containing copies of the values of the first row and column.
This way one avoids boundary artifacts resulting from a naive interpolation that ignores the necessary periodicity at the unit cell boundaries.

The numerical minimization is done with a nonlinear Fletcher-Reeves conjugate gradient descent including Armijo step size control with widening. First, we minimize with respect to $u$ for the given initial guess of $v$ and starting with $u=0$. Then, we minimize over $u$ and $v$ simultaneously.
Note that the energy \eqref{motifimageeq} is non-convex. Nonetheless, applying this first step on numerous experimental and synthetic images with unit cell sizes varying from one to eleven atoms and image sizes from 200 x 200 to 1024 x 1024 have resulted in successful motif extraction. Critical for this success is that the unit cell vector estimate obtained as described in \cref{sec:UnitCellExtr} is quite precise.

\begin{figure}[ht]
    \centering
    \setlength{\imgwidth}{.2075\linewidth}
    \setlength{\tabcolsep}{2pt}
    \begin{tabular}{@{}rcccc@{}}
    &
        {a)}&
        {b)}&
        {c)}&
        {d)}
        \\
        \raisebox{.45\imgwidth}{$f,\textcolor{red}{v_1},\textcolor{ucvecgreen}{v_2},u$}&
        \includegraphics[height=\imgwidth]{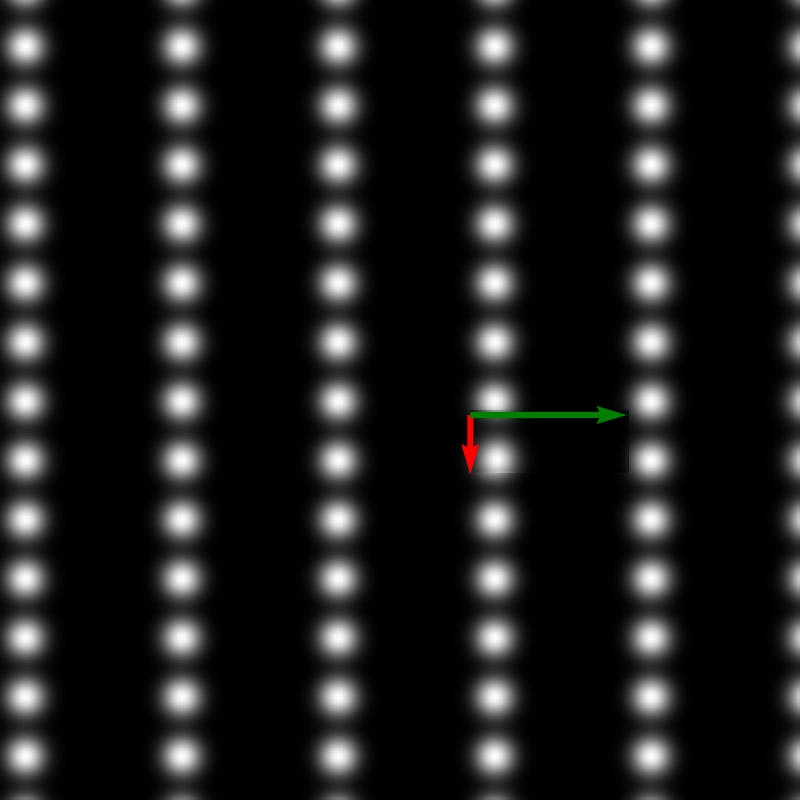}
        &\includegraphics[height=\imgwidth]{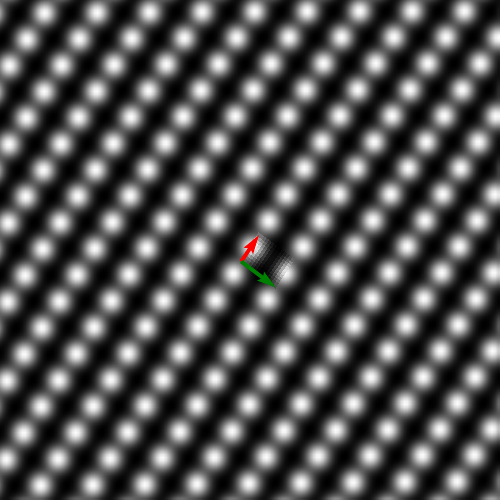}
        &\includegraphics[height=\imgwidth]{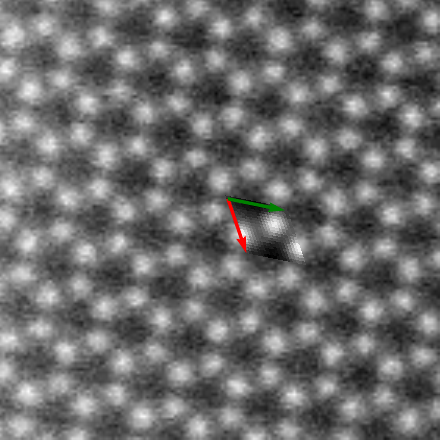}
        &\includegraphics[height=\imgwidth]{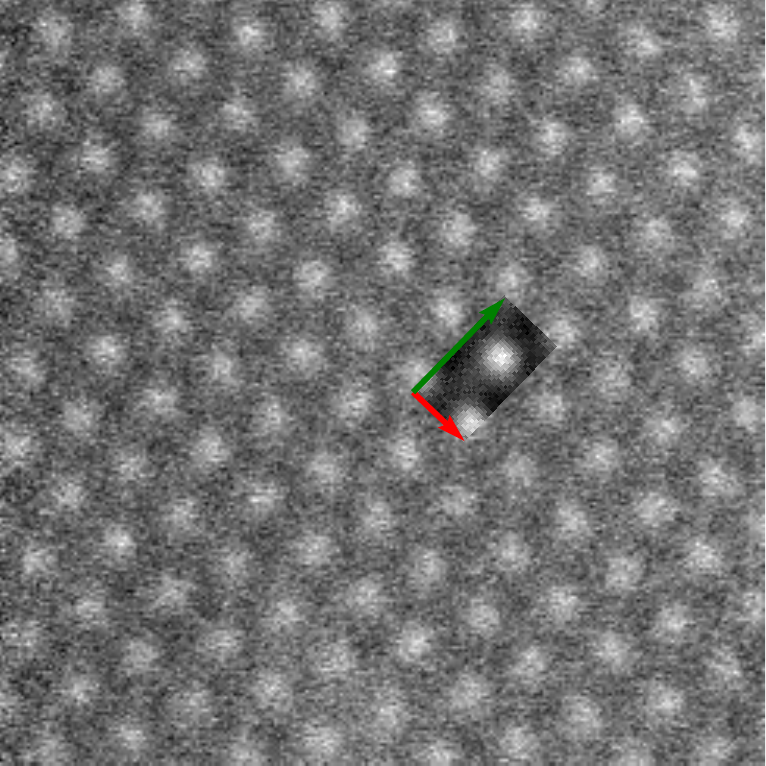}
        \\
        \raisebox{.45\imgwidth}{$u\circ\ProjEuclidToUCCrystal$}&
        \includegraphics[height=\imgwidth]{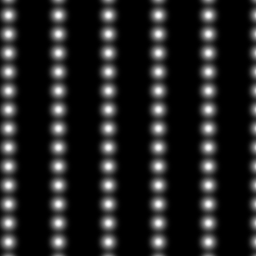}
        &\includegraphics[height=\imgwidth]{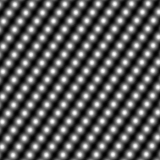}
        &\includegraphics[height=\imgwidth]{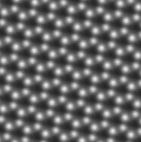}
        &\includegraphics[height=\imgwidth]{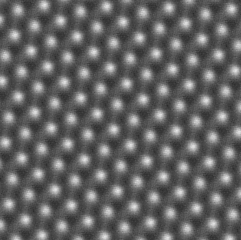}
        \\
        \raisebox{.16\imgwidth}{$u\;$}&
        \includegraphics[height=.37\imgwidth]{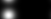}
        &\includegraphics[height=.37\imgwidth]{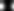}
        &\includegraphics[height=.37\imgwidth]{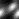}
        &\includegraphics[height=.37\imgwidth]{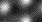}
        \\ %
    \end{tabular}
    \caption{Motif image extraction and input image reconstruction. Images a) and b) are synthetic images while c) and d) are HAADF STEM images of Mg imaged along the [0001] (also shown in \cref{2grains}b)) and [11$\bar{2}$0] zone axis. The first row shows the input images $f$ with the extracted motif image $u$ and the unit cell vectors in red and green drawn on top.
    The second row shows the reconstructed denoised images $u\circ\ProjEuclidToUCCrystal$. The extracted motif images $u$ are shown in the last row. Note that the images in the last row are in crystal space, i.e., the vertical and horizontal axes of the images correspond to $\vee_1$ and $\vee_2$.
    Image b) courtesy of Marvin Poul, MPIE.
    }
    \label{motifimagefig}
\end{figure}

\cref{motifimagefig} shows some examples of the motif image extraction.
Since the resulting motif image $u$ is defined in the crystal space, the vertical and horizontal axes of the images correspond to  $\vee_1$ and $\vee_2$. In order to show the right inclination of these motif images, the change of coordinates $\CoordCrystalToEuclid[v]$ is required.
\cref{motifimagefig} also shows an interesting byproduct of this first step, i.e., the reconstruction of the image $u\circ\ProjEuclidToUCCrystal$. This is an average version of the input image. One of the most important aspects of this reconstructed image is that it is strongly denoised, which is useful in the second step of the motif extraction.

\begin{figure}[ht]
    \label{brickfig}
    \center
    \setlength{\imgwidth}{.4\linewidth}
    \begin{tabular}{@{}m{\imgwidth}m{.36\imgwidth}m{\imgwidth}@{}}
        \includegraphics[height=\imgwidth]{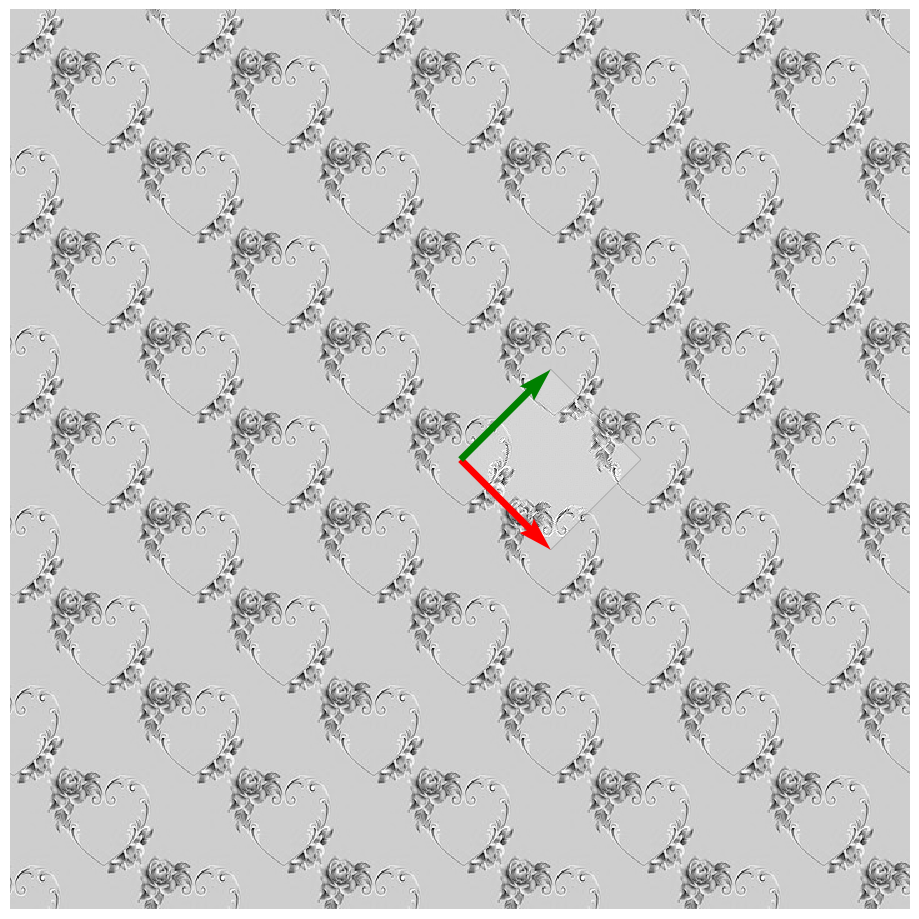}
        &
        \includegraphics[height=.37\imgwidth]{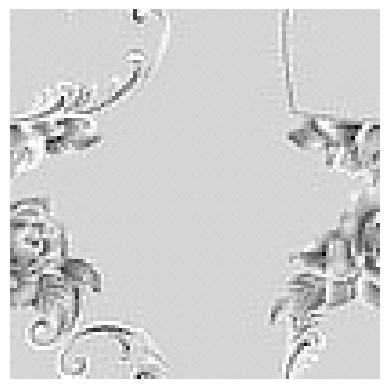}
        &
        \includegraphics[height=\imgwidth]{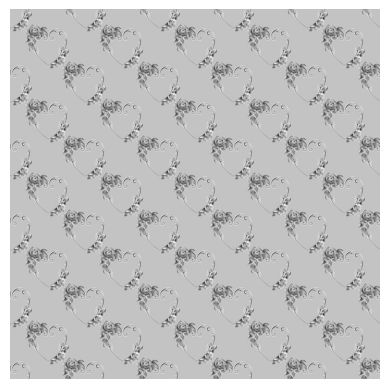}
        \\
        \centering
        $f,\textcolor{red}{v_1},\textcolor{ucvecgreen}{v_2},u$
        &
        \centering
        $u$
        &
        \centering
        $u\circ\ProjEuclidToUCCrystal$
    \end{tabular}
    \caption{Motif extraction example for a periodic non-crystalline image.}
    \label{fig:NonCrystalMotif}
\end{figure}

Furthermore, this motif image extraction is not restricted to micrographs and can be applied to any periodic image. \cref{fig:NonCrystalMotif} exemplifies this on a periodic, non-crystalline image.

\section{Determining the constituent atoms of the motif}
\label{sec:MotifDOFs}
\begin{figure}[ht]
    \center
    \setlength{\imgwidth}{.305\linewidth}
    \begin{tabular}{@{}ccc@{}}
        \includegraphics[width=\imgwidth, trim = 15 15 15 15, clip]{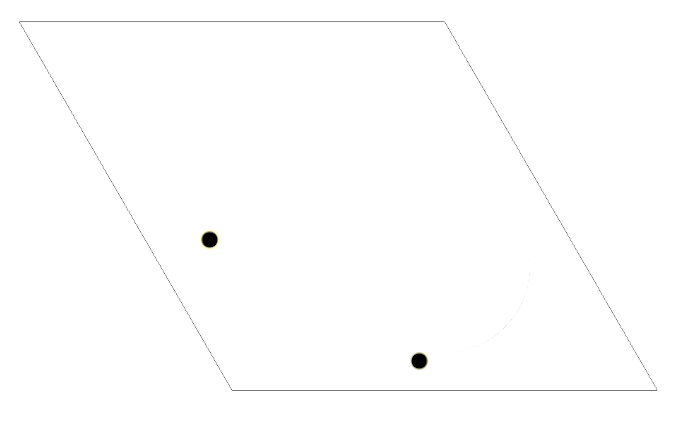}&
        \includegraphics[width=\imgwidth, trim = 15 15 15 15, clip]{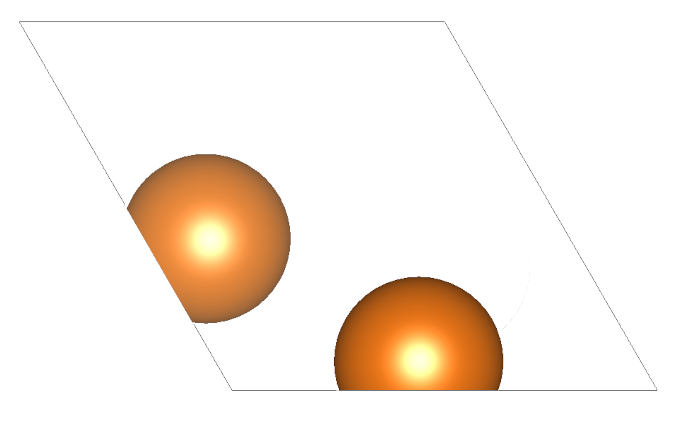}&
        \includegraphics[width=\imgwidth, trim = 15 15 15 15, clip]{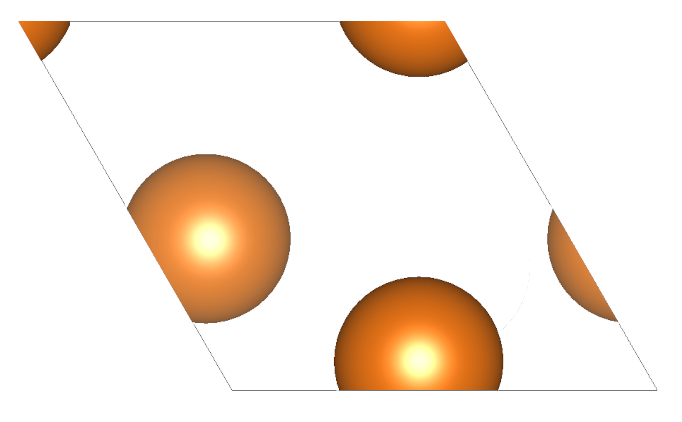}\\
        ${\mu_1,\ldots,\mu_l}$&
        $\motifgaussian$&
        $\varrho$
    \end{tabular}
    \caption{An illustration of the different ingredients of our forward model $\varrho$, which models the motif image inside the unit cell in Euclidean coordinates.}
    \label{varsigfig}
\end{figure}

The second step of the motif extraction is to decompose the motif image into its constituent atoms in terms of their positions and image intensities. For this, we first introduce a forward model that converts the motif into the corresponding atomic resolution image.
To this end, we assume that each atom (or rather atomic column) in the image can be approximated with the following Gaussian bump model
\[
    g[\mu,\sigma,h,r](x)
    =
    h
    \exp
    \left( \frac{- 1}{ 2 ( 1 - r^2 ) }
    \left(
    \left( \frac{x_1-\mu_{1}}{\sigma_{1}} \right)^2
    +
    \left( \frac{x_2-\mu_{2}}{\sigma_{2}} \right)^2
    -
    \frac{2r}{ \sigma_{1} \sigma_{2}} (x_1-\mu_{1} ) (x_2-\mu_{2})
    \right)
    \right).
\]
Here, the parameters of the model / the corresponding Gaussian distribution are:

\begin{tabular}{rl}
  $\mu \in \mathbb{R}^2$ & atom center\\
  $\sigma \in (0,\infty)^2$ & atom width/height / standard deviation along $x_1$ and $x_2$\\
  $h \in [0,\infty) $ & atom intensity / distribution height\\
  $r \in (-1,1) $ & skewness / correlation coefficient
\end{tabular}
\\[1ex]
Assuming that the motif contains $l\in\mathbb{N}$ atomic columns, the image motif can be modeled using the bump model by
\[
\motifgaussian[\motifgaussianparameters](x)
\coloneqq
\sum_{c=1}^l
g[\mu_c,\sigma_c,h_c,r_c](x)
 ,
\]
where $\motifgaussianparameters$ are the bump fitting parameters for each atomic column, i.e.,
\[\motifgaussianparameters \coloneqq (\vb{\mu}_c,\vb{\sigma}_c,h_c,r_c)_{c=1}^l.\]
The unit cell structure $\unitvec_1,\unitvec_2$ and the periodicity still needs to be accounted for. For this, we introduce $\varrho$:
\[
\varrho [\motifgaussian[\motifgaussianparameters],b] (x)
\coloneqq
\sum_{z_1,z_2=-1}^1
\motifgaussian[\motifgaussianparameters]%
(x+z_1 \unitvec_1 +z_2 \unitvec_2 )
+
b\]
Here, $b \in \mathbb{R}$ is the background intensity, which is also handled by $\varrho$.
\cref{varsigfig} illustrates the construction of $\varrho$. Essentially, all atomic columns inside the unit cell are copied along the unit cell vectors to account for the effects of the periodicity.
With $\varrho$, we can finally formulate the energy for the second part of the motif extraction, i.e.,
\begin{equation}
    \label{motifatmseq}
    \begin{split}
        \EMotifAtm[\motifgaussianparameters, b, v] \coloneqq%
        \int_{\Omega} \left(
        f(x) -
        \varrho[\motifgaussian[\motifgaussianparameters], b](\ProjUCEuclid[v](x))
        \right)^2 \dx.
    \end{split}
\end{equation}
$\EMotifAtm$ is minimized over $\motifgaussianparameters, b$, keeping $v$ from the previous step fixed.
Like for $\EMotifIm$, we use a nonlinear Fletcher-Reeves conjugate gradient descent including Armijo step size control with widening for the minimization of $\EMotifAtm$.
An important aspect here for the successful minimization of the highly non-convex function $\EMotifAtm$ is to have a good initial guess for the atomic column centers and widths, i.e., $(\vb{\mu}_c,\vb{\sigma}_c)_{c=1}^l$ in $\motifgaussianparameters$. To find these values, the reconstructed image $u\circ\ProjEuclidToUCCrystal$ from the first step is very beneficial. Since this is a strongly denoised version of the input image $f$, finding all atomic columns is possible with standard bump fitting approaches. The fitted atomic column positions are projected to the unit cell in Euclidean space using $\ProjUCEuclid$. The resulting Euclidean space positions in the unit cell are clustered using the $k$-means algorithm into $l$ clusters, the corresponding cluster centers are suitable initial guess for $(\vb{\mu}_c)_{c=1}^l$. The mean values of $\sigma$, $h$ and $r$ from the simple fit on the reconstructed image are used as initial guess for $(\vb{\sigma}_c,h_c,r_c)_{c=1}^l$.
In Algorithm~\ref{algmotif}, we sketch all steps of the entire motif extraction algorithm.
\IncMargin{1em}
\begin{algorithm}[ht]
	\SetAlgoLined
	\caption{Overview of the full motif extraction algorithm}
	\label{algmotif}
	\KwIn{ periodic image $f:\Omega\to\mathbb{R}$, 
            number of atoms in the primitive unit cell $l \in \mathbb{N}$ 
	;}
	\KwResult{
		\begin{tabular}[t]{@{}l@{}}	
			primitive unit cell vectors $\vee \in \mathbb{R}^{2,2}$,\\
			motif $u :[0,1)^2\rightarrow\mathbb{R}$,  \\
			atomic parameters $\motifgaussianparameters = (\vb{\mu}_c,\vb{\sigma}_c,h_c,r_c)_{c=1}^l$ and $b.$
		\end{tabular}
	}
	\begin{enumerate}[noitemsep, leftmargin=3ex]
		\item $\vee^1 \leftarrow $ unit cell extraction of $f$ (see \cref{sec:UnitCellExtr}) \;
		\item $u^1 \leftarrow$ minimize $u\mapsto\EMotifIm[u,\unitvec^1]$ (see \eqref{motifimageeq}), initial value $u = 0$  \;
		(note $\EMotifIm$ includes periodic interpolation, see \cref{secmotifimg}) \;
		\item $u^2,\vee^2 \leftarrow$ minimize $(u,v)\mapsto\EMotifIm[u,v]$, initial values $u = u^1,\vee = \vee^1$\;
		\item $(\vb{\mu}_c^0,\vb{\sigma}_c^0,h_c^0,r_c^0)_{c=1}^N$ $\leftarrow$ standard bump fitting of $ (u^2 \circ \ProjUCEuclid[\vee^2])$\\
		$(\vb{\mu}_c^1)_{c=1}^l \leftarrow$ $k$-means of $(\ProjUCEuclid[v^2](\vb{\mu}_c^0))_{c=1}^N$ with $l$ clusters\\
		$(\vb{\sigma}_c^1,h_c^1,r_c^1)_{c=1}^l \leftarrow$ mean of $(\vb{\sigma}_c^0,h_c^0,r_c^0)_{c=1}^N$ over $c$\;
		\item $\Theta^2,b^1 \leftarrow$ minimize $(\Theta,b)\mapsto\EMotifAtm[\Theta,b,\vee^2]$,\\\hspace{9.75ex}initial values $\Theta=(\vb{\mu}_c^1,\vb{\sigma}_c^1,h_c^0,r_c^0)_{c=1}^l,b = 0$ (see \eqref{motifatmseq})\;
	\end{enumerate}
	\vspace{-.5\baselineskip}
	\KwRet{$v^2$, $u^2$, $\Theta^2$, $b^1$}
\end{algorithm}

\begin{figure}[p]
    \center
    \setlength{\imgwidth}{.4\linewidth}
    \begin{tabular}{@{}m{.1\imgwidth}m{\imgwidth}m{.5\imgwidth}@{}}
        a)&
        \includegraphics[page=4,width=\linewidth]{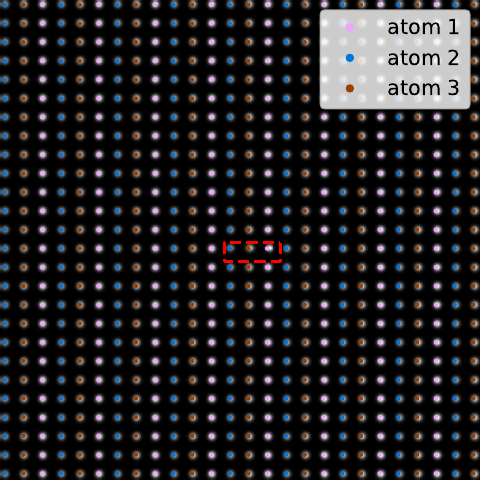}&
        \includegraphics[page=4,trim = 3.6cm 3.6cm 3.3cm 3.9cm, clip,width=\linewidth]{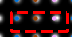}\\
        b)&
        \includegraphics[page=4,width=\linewidth]{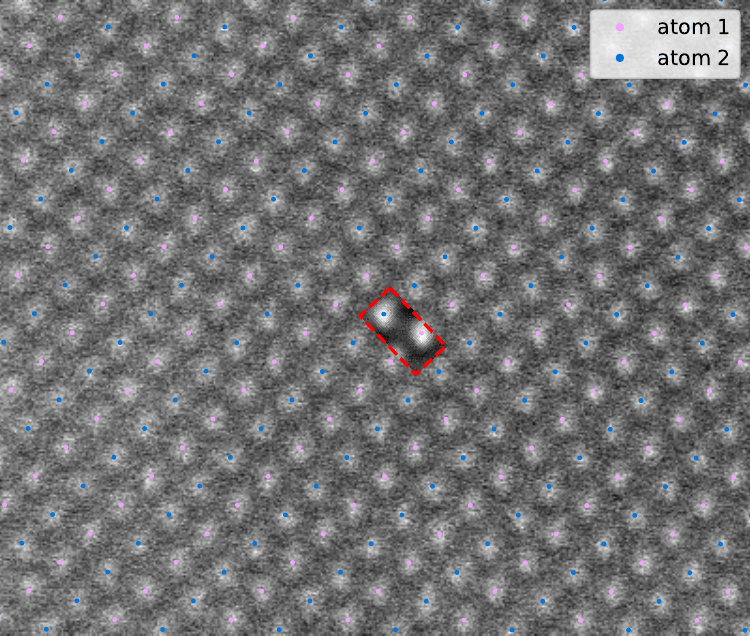}&
        \includegraphics[page=4,trim = 6cm 4.3cm 5.1cm 4.8cm, clip,width=\linewidth]{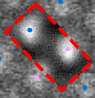}\\
        c)&
        \includegraphics[page=4,width=\linewidth]{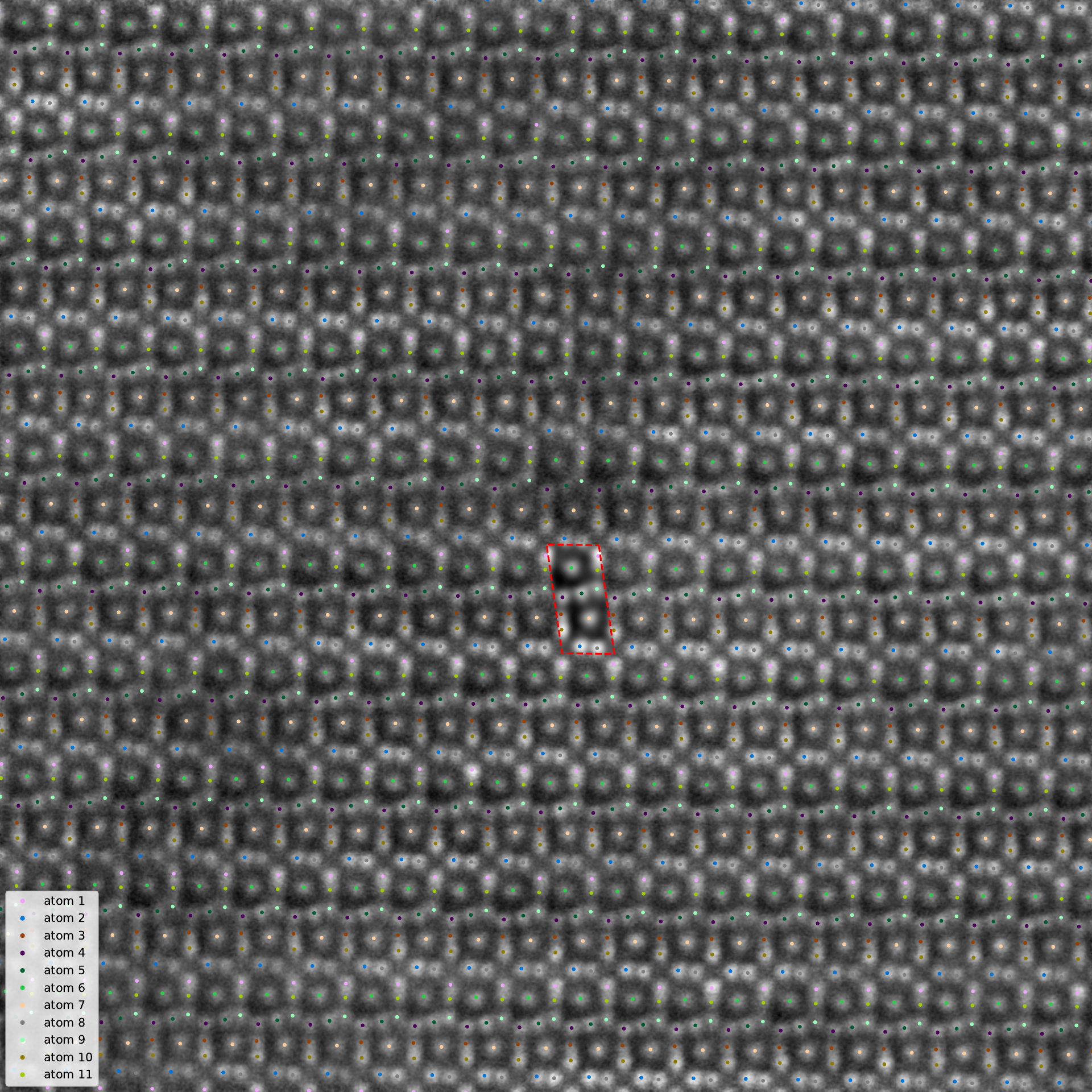}&
        \includegraphics[page=4,trim= 16cm 12.8cm 14cm 16cm,clip,width=\linewidth]{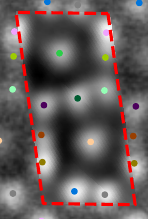}
    \end{tabular}
    \caption{Results of the full motif extraction on a) a synthetic lattice with 3 atoms in the unit cell, b) an experimental HAADF STEM image of a Mg crystal (\cref{motifimagefig}.$d$) and c) Nb\textsubscript{6.4}Co\textsubscript{6.6} imaged along the [11$\bar{2}$0] axis zone \citep{Luo_Xie_Zhang_2023}.}
    \label{motifexfig}
\end{figure}

\cref{motifexfig} exemplifies the motif extraction results from the second step. The red bordered rectangle shows the unit cell, inside the rectangle the motif image $u$ from the first step is displayed transformed to Euclidean coordinates.

\begin{figure}[ht]
    \center
    \setlength{\imgwidth}{.32\linewidth}
    \setlength{\tabcolsep}{3pt}
    \begin{tabular}{@{}ccc@{}}
        \includegraphics[page=1,width=\imgwidth]{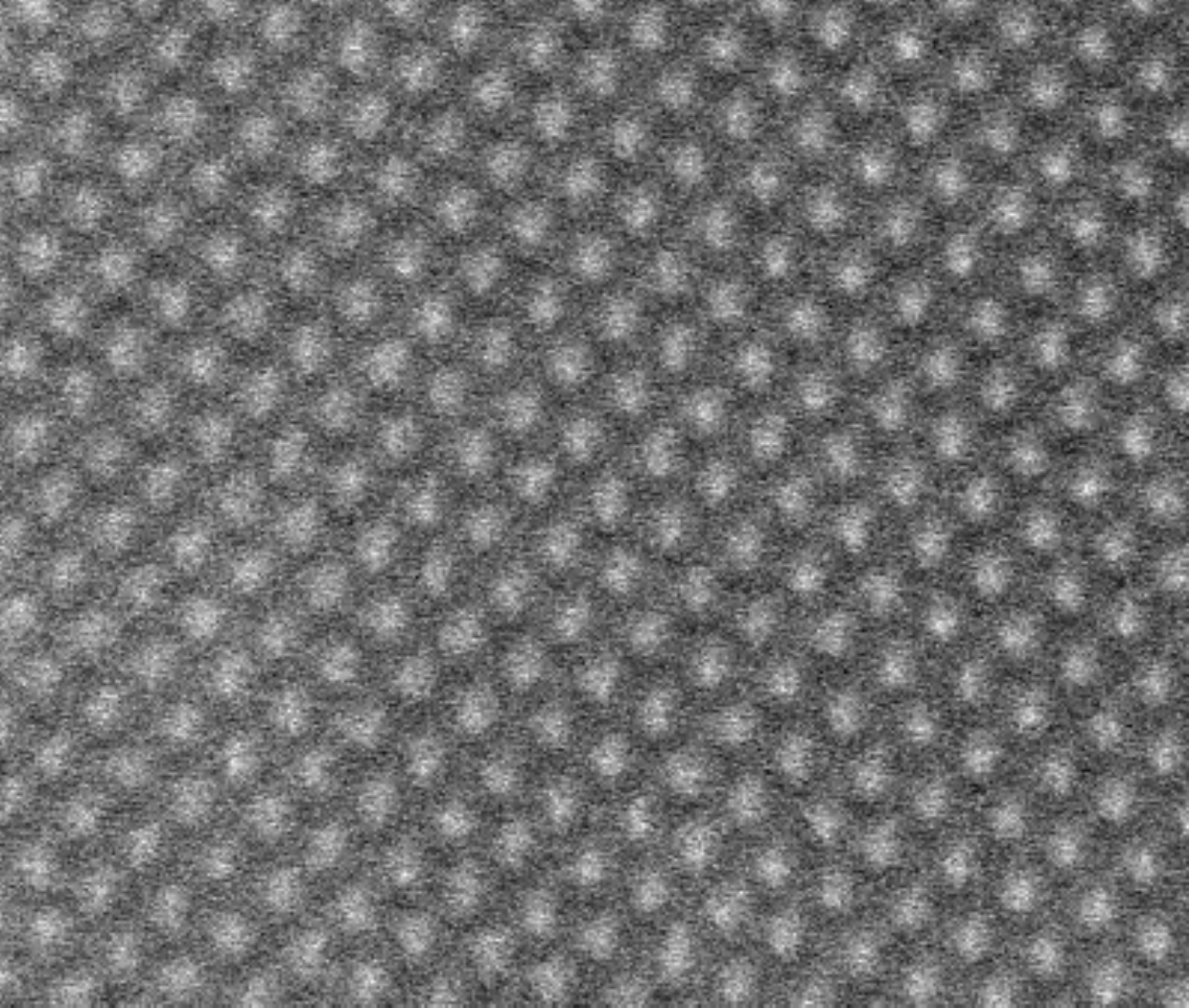}
        &
        \includegraphics[width=\imgwidth]{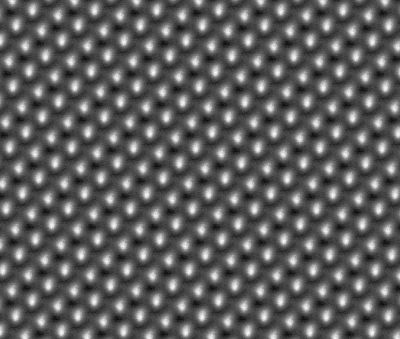}
        &
        \includegraphics[page=3,width=\imgwidth]{figures/tb_1051_right_motif_results.pdf}
        \\
        \includegraphics[page=1,width=\imgwidth]{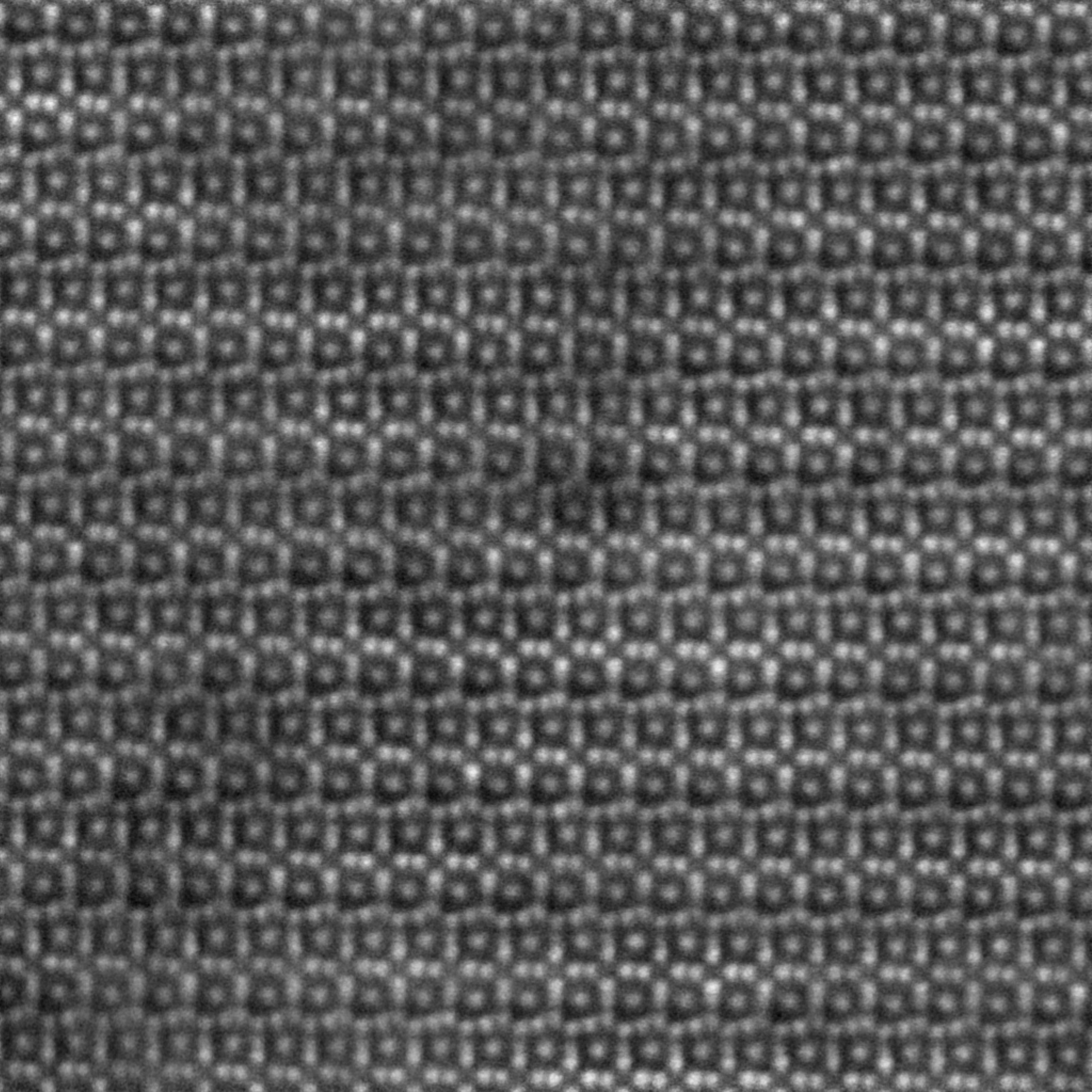}
        &
        \includegraphics[width=\imgwidth]{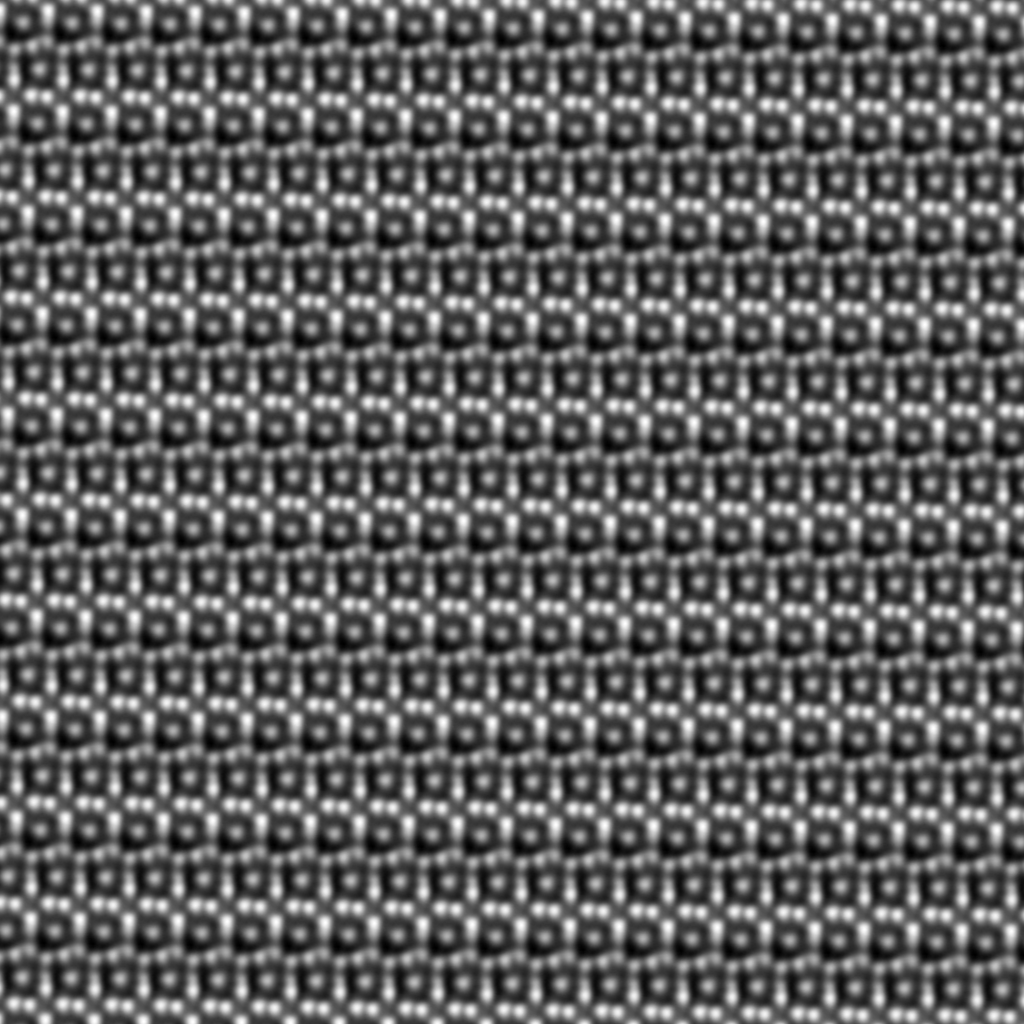}
        &
        \includegraphics[page=3,width=\imgwidth]{figures/1552_110average_motif_results.pdf}
        \\
        $f$
        &
        $u\circ\ProjEuclidToUCCrystal$
        &
        $\varrho[\motifgaussian[\motifgaussianparameters],b](\ProjUCEuclid)$
    \end{tabular}
    \caption{The input image $f$, the reconstructed image from the first step $u\circ\ProjEuclidToUCCrystal$ and the model reconstruction $\varrho[\motifgaussian[\motifgaussianparameters],b]\circ\ProjUCEuclid$ from the second step for the two STEM images from \cref{motifexfig}.}
    \label{fig:reconstruction}
\end{figure}

The forward model combined with the projection to the unit cell in Euclidean coordinates, i.e., $\varrho[\motifgaussian[\motifgaussianparameters],b]\circ\ProjUCEuclid$, also serves as a model reconstruction of the input image. \cref{fig:reconstruction} shows the input image, the image reconstruction from the first step and the model reconstruction for the Mg data and Nb\textsubscript{7}Co\textsubscript{6}, respectively.

\section[A motif extraction application in complex unit cells of topologically close-packed \textmu phases]{A motif extraction application in complex unit cells of topologically close-packed $\mu$ phases}
\label{application}

\begin{figure}[ht]
    \centering
    \setlength{\imgwidth}{.4\linewidth}
    \begin{tabular}{@{}m{\imgwidth}m{\imgwidth}m{.2\imgwidth}@{}}
        \includegraphics[width=\imgwidth,trim=0 0 6.7cm 0,clip%
        ]{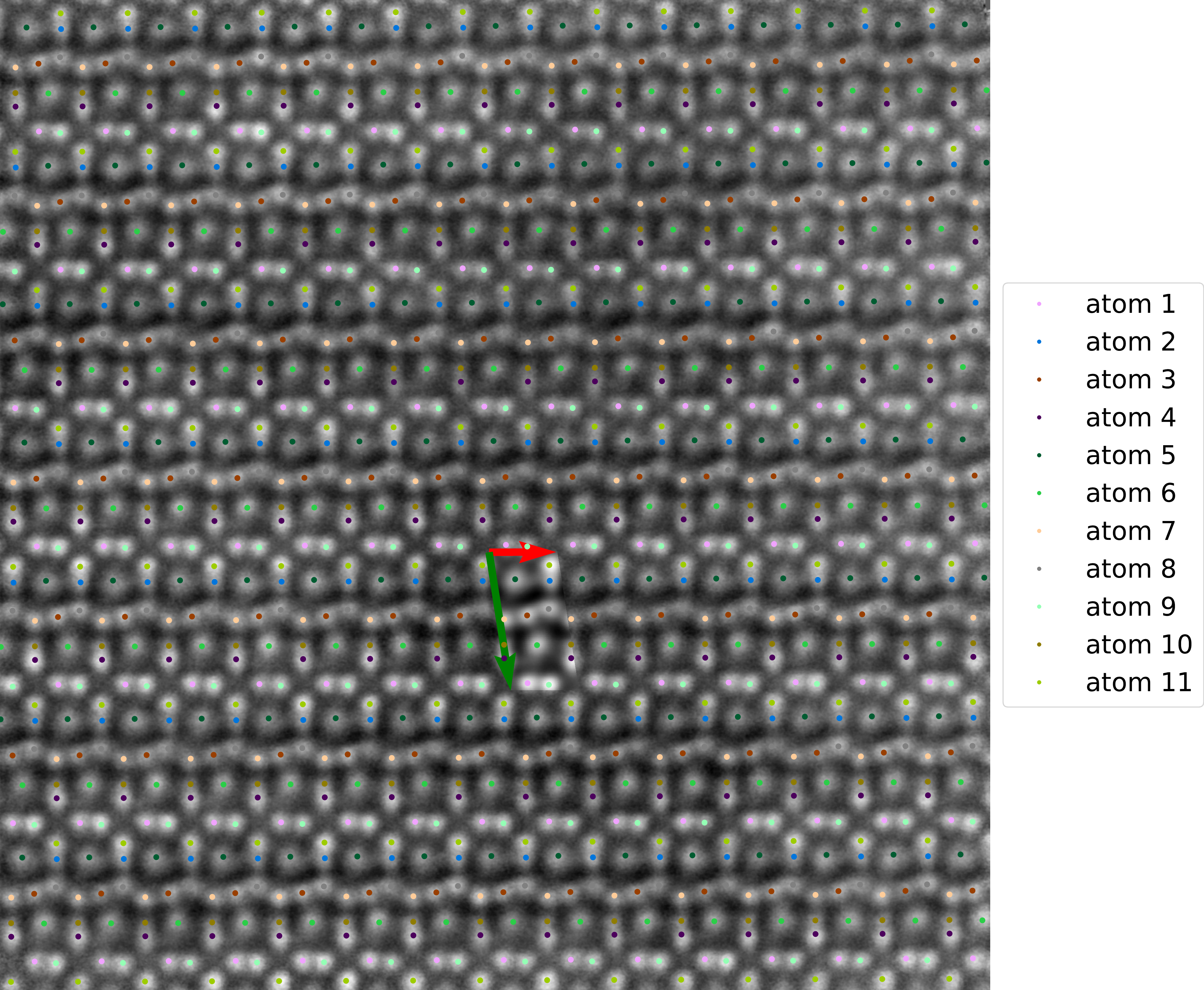}
        &
        \includegraphics[width=\imgwidth,trim=0 0 6.7cm 0,clip%
        ]{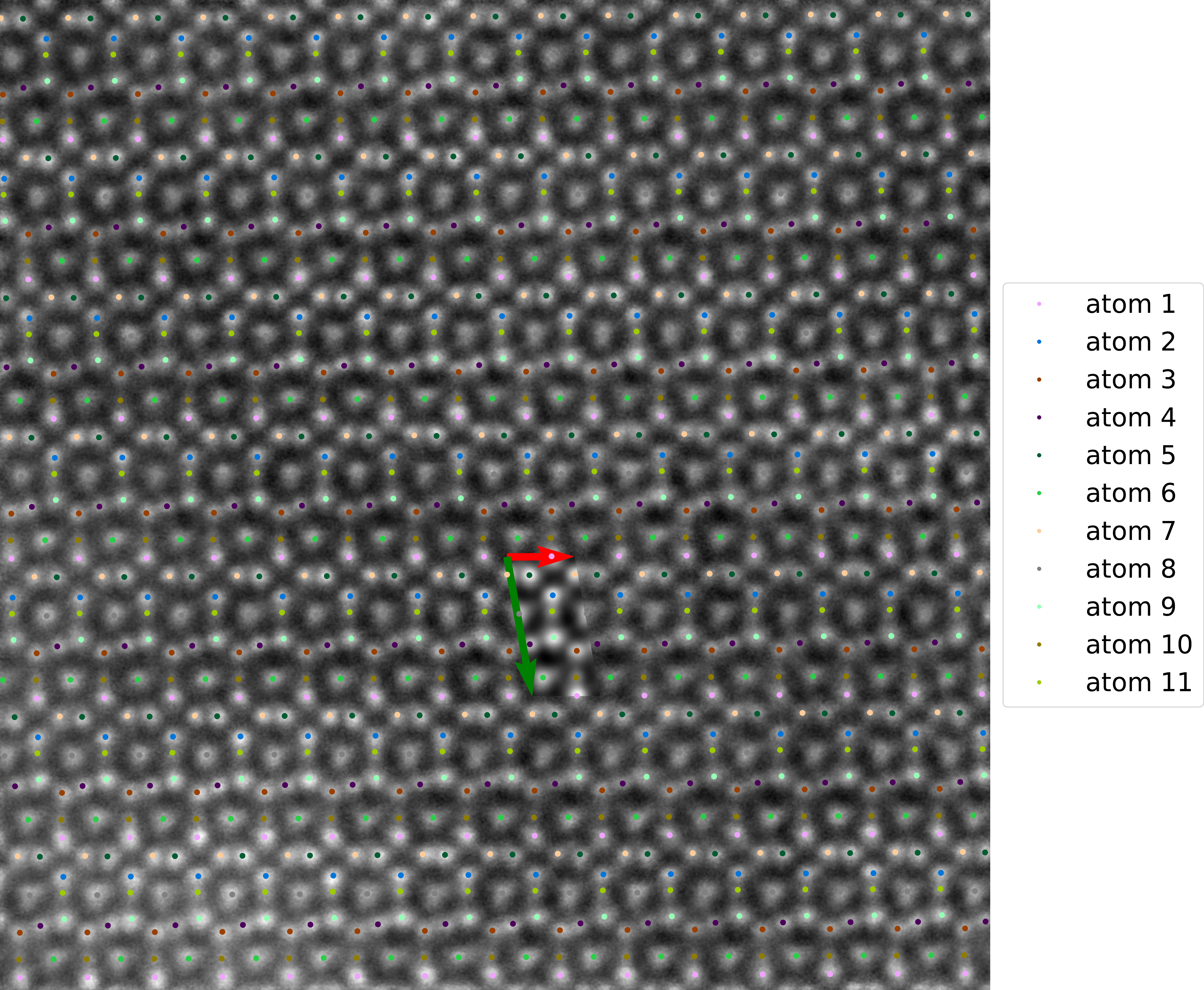}
        &
        \includegraphics[width=.3\imgwidth,trim=33cm 9cm 0 9.5cm,clip,scale=.4%
        ]{figures/NbCo_1659_motifonimg1.pdf}
        \\
        \center{Nb\textsubscript{6.4}Co\textsubscript{6.6}}
        &
        \center{Nb\textsubscript{7}Co\textsubscript{6}}
        &
        \\[1ex]
    \end{tabular}
    \caption{Results of the motif extraction of complicated motif with 11 atoms in the 2D projection of the primitive unit cell  along the [11$\bar{2}$0] zone axis.}
    \label{nbcofig}
\end{figure}

Finally, we showcase the application of motif extraction to a complex Nb\textsubscript{6}Co\textsubscript{7} $\mu$-phase unit cell containing 39 atoms.
It has been reported that this phase has a wide range in chemical composition, which can lead to fundamentally different mechanical properties.
For this, our method was used to extract the primitive cells of atomic resolution STEM images of Nb\textsubscript{6.4}Co\textsubscript{6.6} and Nb\textsubscript{7}Co\textsubscript{6} projected along the [11$\bar{2}$0] zone axis. As shown in \cref{nbcofig}, instead of the hexagonal unit cell containing 39 atoms, a primitive cell containing 13 atoms in 3D and 11 on the 2D projection was considered. 
\begin{figure}
    \centering
        \setlength{\imgwidth}{.32\linewidth}
        \includegraphics[width=\imgwidth]{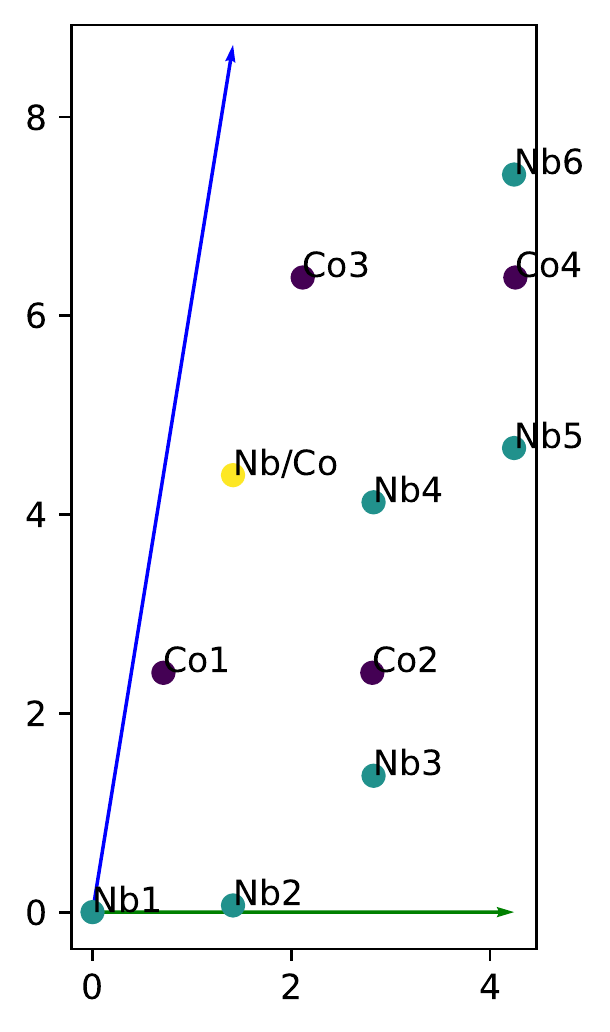}
        \caption{Reference unit cell of Nb\textsubscript{6.4}Co\textsubscript{6.6} projected along the [11$\bar{2}$0] zone axis with labels for all atomic columns.}
    \label{figNbColabeling}
\end{figure} 
In particular the atomic columns \textbf{Nb4}, \textbf{Nb/Co} and \textbf{Nb5} constitute the so-called triple layer, whose interplanar distance controls the plastic behavior of the crystal \citep{Luo_Xie_Zhang_2023}. \cref{figNbColabeling} shows the projected reference unit cell with labelled atomic columns for Nb\textsubscript{6.4}Co\textsubscript{6.6} for the case that $v_1$ is aligned to the $x_1$-axis. The interplanar distance of the triple layer is given by
\[d_t  = \mathbf{Nb5}(x_2) - \mathbf{Nb/Co}(x_2) = \mathbf{Nb/Co}(x_2) - \mathbf{Nb4}(x_2).\]
Here, $\mathbf{A}(x_2)$ denotes the vertical position of the atom column $\mathbf{A}$.
Since the motif extraction determines the average atomic column positions in the motif for an entire images, we are able to perform a statistical analysis on this interplanar distance.
For the formula for $d_t$ to apply, the extracted motif and the unit cell are rotated such that $v_1$ is in direction to the $x_1$ axis. Moreover, the atomic columns \textbf{Nb4}, \textbf{Nb/Co} and \textbf{Nb5} have to be identified in the extracted motif. For the results shown here, the necessary motif labelling was done manually.

For the Nb\textsubscript{7}Co\textsubscript{6} image shown in \cref{nbcofig}, the triple layer spacing was determined as \SI[parse-numbers=false]{42.85 \pm 0.35}{\pico\metre}.
For the Nb\textsubscript{6.4}Co\textsubscript{6.6} image shown in \cref{nbcofig}, it was determined as \SI[parse-numbers=false]{32.6 \pm 8.4}{\pico\metre}. Due to the high standard deviation of the latter image, 8 more images have been acquired for the Nb\textsubscript{6.4}Co\textsubscript{6.6} sample, which allowed us to apply the proposed approach to analyze the statistical scattering of the value due to, e.g., different areas of the sample. The results are
\SI[parse-numbers=false]{34.55 \pm 0.32}{\pico\metre},
\SI[parse-numbers=false]{34 \pm 0.3}{\pico\metre},
\SI[parse-numbers=false]{31.22 \pm 0.64}{\pico\metre},
\SI[parse-numbers=false]{31.46 \pm 0.78}{\pico\metre},
\SI[parse-numbers=false]{32.66 \pm 0.13}{\pico\metre},
\SI[parse-numbers=false]{32.65 \pm 3.63}{\pico\metre},
\SI[parse-numbers=false]{27.51 \pm 3.69}{\pico\metre} and
\SI[parse-numbers=false]{30.39 \pm 7.66}{\pico\metre}.
The mean and the standard variation of these 9 means lead to a triple layer spacing of \SI[parse-numbers=false]{31.89 \pm 1.98}{\pico\metre} for the Nb\textsubscript{6.4}Co\textsubscript{6.6} sample.
Therefore, we can conclude that there is a statistical difference between the measured triple layer spacing between Nb\textsubscript{7}Co\textsubscript{6} and Nb\textsubscript{6.4}Co\textsubscript{6.6}.

This statistical analysis was used in \citep{Luo_Xie_Zhang_2023}. There, it was concluded that the DFT calculations are consistent with these results measured from the experimental HAADF-STEM images.

\section{Conclusion}
We proposed a novel framework for the extraction of the motif from a crystalline image at atomic scale. The framework builds on an automatic extraction of the unit cell vectors as starting point. The motif extraction then is done in two steps. The first step extracts the motif in terms of an image, the second step determines the constituent atoms of the motif using a periodized bump fit with a general 2D Gaussian distribution. The parameters of this model are the bump position, width in $x_1$ and $x_2$ direction, the correlation between $x_1$ and $x_2$, the Gaussian height and the background intensity. %
Denoised and model images of the input image are byproducts of the method.
The method was successfully applied to various synthetic and experimental images. %

\section*{Acknowledgement}
    This work was supported by the German research foundation (DFG) within the Collaborative Research Centre SFB 1394 \enquote{Structural and Chemical Atomic Complexity—From Defect Phase Diagrams to Materials Properties} (Project ID 409476157), A.S.A.A. and B.B. in Project A04, S.Z. in Project B01.

\bibliographystyle{elsarticle-num-names}
\bibliography{abpaper1_references}
\end{document}